# Broadband dual-comb hyperspectral imaging and adaptable spectroscopy with programmable frequency combs

**FABRIZIO R. GIORGETTA[1,2], JEAN-DANIEL DESCHÊNES[3], RICHARD L. LIEBER[4,4A], IAN CODDINGTON[1], NATHAN R. NEWBURY[1] AND ESTHER BAUMANN[1,2]**

[1]*National Institute of Standards and Technology, Communications Technology Division, Boulder, CO 80305,* [2]*University of Colorado, Boulder, Department of Physics, Boulder, CO 80309,* [3]*Octosig Consulting, Québec City, Québec, Canada,* [4]*Shirley Ryan Ability Lab Chicago,* [4A]*Hines VA Medical Center Chicago*

baumann@nist.gov

**Abstract**:

We explore the advantages of a free-form dual-comb spectroscopy (DCS) platform based on time-programmable frequency combs for real-time, penalty-free apodized scanning. In traditional DCS, the fundamental spectral resolution, which equals the comb repetition rate, can be excessively fine for many applications. While the fine resolution is not itself problematic, it comes with the penalty of excess acquisition time. Post-processing apodization (windowing) can be applied to tailor the resolution to the sample, but only with a deadtime penalty proportional to the degree of apodization. The excess acquisition time remains. With free-form DCS, this deadtime is avoided by programming a real-time apodization pattern that dynamically reverses the pulse periods between the dual frequency combs. In this way, one can tailor the spectrometer's resolution and update rate to different applications without penalty. We show operation of a free-form DCS system where the spectral resolution is varied from the intrinsic fine resolution of 160 MHz up to 822 GHz by applying tailored real-time apodization. Because there is no deadtime penalty, the spectral signal-to-noise ratio increases linearly with resolution by 5000x over this range, as opposed to the square root increase observed for post-processing apodization in traditional DCS. We explore the flexibility to change resolution and update rate to perform hyperspectral imaging at slow camera frame rates, where the penalty-free apodization allows for optimal use of each frame. We obtain dual-comb hyperspectral movies at a 20 Hz spectrum update rate with broad optical spectral coverage of over 10 THz.

## 1. Introduction

Dual-comb spectroscopy (DCS) probes many spectral signatures simultaneously with the broad spectrum of an optical frequency comb. Resolved spectra are generated by interference with a second comb and without mechanical interferometer or dispersive elements. With fast update rates, no moving parts, and a coherent light source, DCS has often been promoted as a novel alternative to the classical Fourier transform infrared spectrometer (FTIR) workhorse. Dual-comb spectrometers however still do not come with the versatility of FTIRs, particularly when it comes to adapting the spectral resolution and update rate to match the sample. Over the years DCS has been applied to multiple spectroscopic applications including high-precision molecular spectroscopy, open path green-house gas monitoring, breath analysis, fast molecular reaction measurements, and non-linear spectroscopy [1–3]. What has not changed is the traditional operation scheme; two combs with rigidly fixed repetition rates, $\sim f_r$, are used to interrogate the sample. A pre-set repetition rate of the two underlying combs dictates the spectral resolution, $f_r$, and update rate of the instrument. Unfortunately, when a resolution as fine as $f_r$ is not needed, the excess resolution restricts the achievable signal-to-noise ratio by imposing an overly slow update rate (SNR, here defined for a one-second magnitude spectrum as the ratio between its peak level and its level in a dark spectral region, see appendix). Thus, there is a need to introduce flexibility into the DCS probing paradigm. There are approaches that offer some $f_r$ tuning, such as electro-optic combs [4] or optical cavity tuning [5–7]. Here, we focus on modelocked lasers, which offer both a broad spectral coverage and, in conjunction with our free-form DCS platform, dynamic and fast $f_r$ tuning.

The solution to optimizing resolution is real-time apodization, which was first proposed and implemented for DCS by translating an intracavity mirror using analog electronics in 2005 [8], though the translation was not well controlled making coherent averaging of signals difficult. In real-time apodization, the pulse period difference of the two comb sources is actively inverted repeatedly with minimal dead time. The dual-comb spectrometer then records on a fraction of an interferogram around zero-pulse delay before starting the next one, resulting in a loss in spectral resolution but corresponding gain in update rate and SNR. DCS with real-time apodization operates in close analogy to an FTIR, where the physical scanning range is changed by moving parts [9,10]. One can also mimic this operation in traditional DCS by apodizing in post-processing. However, such post-processing apodization only trades spectral resolution for deadtime; the fundamental spectrum update rate is unaltered.

Because of challenges of controlling the frequency combs, DCS real-time apodization has not been further pursued until recently. Fortunately, recent advances in digital phase locking allow one to precisely program real-time apodization by digitally modulating the set points of the phase locks stabilizing the optical frequency comb [11]. Here we demonstrate real-time apodization with a single free-form dual-comb platform, which allows one to arbitrarily set the pulse time offsets between the combs with high precision [12,13]. While real-time apodization has been used to increase measurement update rates [8,11], here we use free-form DCS to maintain coherent, attosecond control over the combs during the apodization and we study the relation between spectral resolution, SNR, covered spectral bandwidth and update rate.

We implement the free-form DCS using a robust, programmable, Er-fiber dual-comb platform with repetition rates of $f_r$=160 MHz. This low repetition rates offer ease of spectral tuning and spectral broadening through nonlinear processes due to higher pulse energies along with fine intrinsic spectral resolution. The fine resolution also captures narrow spectral features, which could 'slip' between comb teeth and go undetected for higher-$f_r$ DCS systems. By use of real-time apodization, the resolution can be matched to the experimental requirements and is varied from 160 MHz to 822 GHz with no dead-time penalty thus fully replicating the SNR of a DCS constructed with combs at equivalent repetition rates.

We also show that by decoupling the spectral resolution from the comb repetition rate, we can adapt the measurement rate to slow detectors such as a camera focal plane array (FPA). Hyperspectral DCS imaging has recently been performed with electro-optic comb [14–18], but with free-form DCS and real-time apodization it is now possible to harness the far broader spectral coverage possible with mode-locked lasers. Compared to single element photodetectors, infrared focal plane array cameras have update rates of at most a few kHz. Real-time apodization allows sufficient hyperspectral movie update rates to capture dynamic processes while covering unprecedented optical bandwidths. Here we show a 20 Hz hyperspectral movie update rate while interrogating a >10 THz optical bandwidth exceeding previous dual-comb hyperspectral imaging bandwidths by orders of magnitude.

## 2. Real-time apodization on a free-form DCS platform

In DCS, the acquired interferogram is the product of the electric fields of two optical frequency combs at an array of relative pulse delays $\tau_{RPD}$. Figure 1a shows an example interferogram as a function of $\tau_{RPD}$. The interferogram depicts the time domain response of methane convolved with the time domain signal of the comb spectrum centered at 184 THz. The center part of the interferogram, around $\tau_{RPD}=0$ is called centerburst and contains information of the combs' spectral envelope and any broad spectral signature of the sample. The tail of the interferogram contains the free induction decay signature corresponding to the narrow molecular features of methane. In traditional DCS this interferogram is sampled monotonically covering relative pulse delays ranging from $-1/(2f_r)$ to $1/(2f_r)$, indicated by the grey dashed line in Fig. 1b, yielding spectra at the intrinsic spectral resolution of $f_r$. When the acquired spectrum only contains features significantly broader than $f_r$, there is no detectable signal towards the wings of the interferogram and only noise is acquired at large $|\tau_{RPD}|$.

In contrast, in real-time apodization only a reduced portion, $|\tau_{RPD}| \leq \tau_{apod}/2$, of the interferogram is sampled, indicated by the yellow arrow in Fig. 1b. This tailored triangular scanning of $\tau_{RPD}$ changes the spectral resolution from $f_r$ to $f_{res} = 1/\tau_{apod}$, and the spectral SNR increases $\propto f_{res}/f_r$. If $f_{res}$ is matched to the sample's spectral feature width, the SNR improves without loss of spectral information [19]. The measurement acquisition time per spectrum is $a\tau_{apod}$, where $a$ is the programmable slew rate of the free-form dual-comb platform and denotes how fast $\tau_{RPD}$ changes per second. Note that this same apodization can be carried out in post-processing by applying a window as indicated by the solid green arrow in Fig. 1b, but at the cost of a fractional deadtime of $1 - f_r/f_{res}$ with the spectral SNR only increasing $\propto \sqrt{f_{res}/f_r}$.

Figure 1c shows our measurement setup schematic. We implement real-time apodized scanning on the phase stabilized free-form dual-comb platform [13] using two Er-fiber mode locked laser with $f_r = 160$ MHz. The combs are phase stabilized by locking each $f_{ceo}$ signal and by locking one comb tooth with $f_{opt}$ to a common narrow linewidth single mode CW laser at frequency $\nu_{cw} \approx$ 191.56 Hz. To implement the continuous triangular $\tau_{RPD}$ scanning pattern for a given $\tau_{apod}$, a corresponding dynamic phase profile is calculated in a digital signal processor and applied to the lock points $f_{ceo}$, $f_{opt}$ of the digital locking electronics (see appendix).

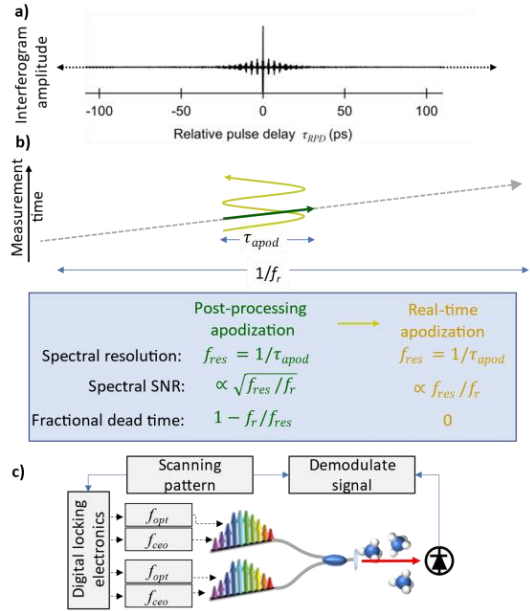

*Figure 1: Real-time apodization DCS overview. a) Optical interferogram centerburst showing methane free-induction decay as a function of relative pulse delay. b) Pulse delay scanning pattern. Grey dashed line: traditional DCS with fixed comb repetition rate. Green solid line: post-processing apodization, where only the centerburst is analyzed but the scan covers the same range as traditional DCS. Yellow line: real-time apodization. The bottom panel compares post-processing apodization to real-time apodization. c) Schematic of free-form DCS setup.*

## 3. Tailor spectral resolution and update rate with real-time apodization

Here, we experimentally analyze SNR and update rate for different real-time apodization scans by measuring narrow methane and broad isoprene absorption features. Figure 2a shows the measurement setup. The combs are spectrally broadened and filtered to cover >10 THz, encompassing the Q and R branches of the methane $^{12}CH_4$, $2\upsilon_3$ (0020 F2) transitions as well as the broad isoprene overtone absorption feature around 184 THz. The combined comb light is collimated and sequentially probes a 75 cm long cell filled with 40% methane, in air, at 60 hPa and a 6 cm long cell filled with 100% isoprene at 1100 hPa. The dual-comb spectrum through each of these cells is shown on the right panel of Figure 2a independently.

Figure 2b shows real-time apodized interferograms for relative pulse delay windows, $\tau_{apod}$, ranging from the maximum $1/f_r = 6.25$ ns down to consecutively shorter apodization windows of 389 ps and 22 ps, correspondingly degrading the spectral resolution from 160 MHz to 2.57 GHz and 45 GHz. At a DCS slew rate $a = 33$ ns/s, the measurement time per spectrum drops from $T_{apod} = 192$ ms to 12 ms and 0.749 ms (taking into account some scan turnaround time, see appendix). This slew rate is limited by the dynamic range and speed of the comb actuators, in our case an intracavity piezo electric transducer and current modulation of a pump laser diode (see appendix).

Figure 2c shows the corresponding normalized transmission spectra averaged over 766 ms. Methane shows many distinct narrow spectral absorption features, and isoprene, a heavier molecular compound, has a broad overall structure lacking narrow features at atmospheric pressure. The broadband isoprene feature is observed at full fidelity for all resolutions, whereas the ~1 GHz wide methane lines are broadened and become less distinct at coarser spectral resolutions. This is because, akin to FTIR, apodization introduces an instrument lineshape for spectral resolutions larger than the spectral feature width. This spectral lineshape is the Fourier transform of the effective square-top apodization window in the time domain.

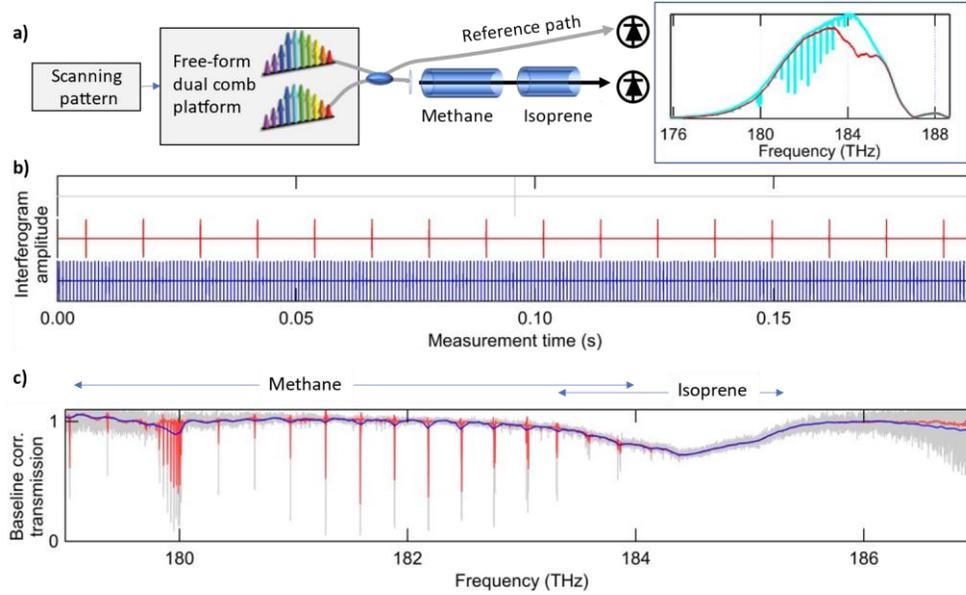

Figure 2: Experimental real-time apodization. a) Setup. A triangular scanning pattern sets the maximum relative pulse delay, or apodization window, between the two combs. The combined light probes methane and isoprene in cells, as well as a reference path. The measured DCS spectrum is shown on the right for just methane (blue) and just isoprene (red). b) Measured real-time apodized interferogram for $T_{apod}$ = 192 ms (top), 12 ms (middle), and 749 µs (bottom). Centerbursts appear as vertical lines at this scale. The shorter $T_{apod}$, the more interferograms are measured, improving the SNR at the cost of spectral resolution. c) Corresponding transmission spectra, normalized with the reference path spectra, showing methane and isoprene features. Each is averaged over 766 ms. As the real-time apodization window shrinks, one observes the expected increase in SNR and loss of spectral resolution, particularly evident for the ~1 GHz wide methane features.

For the $a$ =33 ns/s slew rate used here, a single, unapodized measurement at the full 160-MHz resolution takes 192 ms. Figure 3a shows a real-time apodization measurement taken at a much shorter 12-ms acquisition time with a corresponding resolution of $f_{res}$ =2.57 GHz. This spectrum is fit to a model consisting of methane and isoprene [20,21] and a cubic polynomial baseline. The methane model is calculated at fine resolution (here we use 160 MHz) using Voigt lineshapes and HITRAN 2008 line shape parameters and converted to 2.57 GHz resolution by calculating its inverse Fourier transform, applying the 389 ps apodization window, and finally Fourier transforming back into the spectral domain. (See appendix for a detailed description of the analysis). As shown in Fig. 3b, 128 such measurements are acquired over 1.53 seconds. We can contrast this to traditional DCS followed by post-processing apodization to achieve the same $f_{res}$ =2.57 GHz. In that case, the acquisition time is fixed at 192 ms and only 8 measurements are acquired over 1.53 seconds.

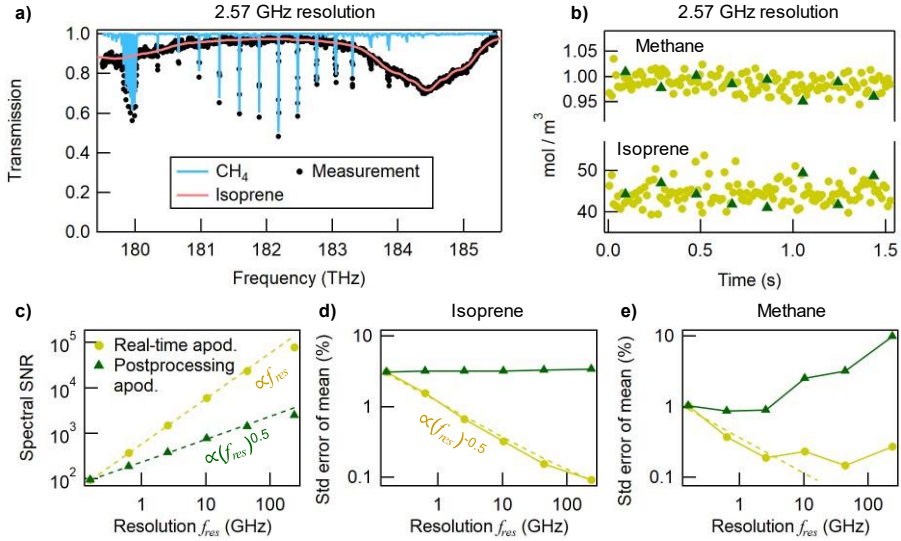

*Figure 3: Spectral analysis. a) Normalized transmission spectrum and fit for a single 2.57 GHz resolution measurement showing the measured spectrum after correction with a cubic polynomial baseline and the fitted methane and isoprene spectra. b) Fitted number density for methane (75 cm long cell) and isoprene (6 cm long cell) at $f_{res}$ =2.57 GHz acquired over 1.53 seconds. Yellow circles denote real-time apodization, green triangles post-processing apodization. c) Spectral SNR for real-time versus post-processing apodization for different spectral resolutions and a fixed 1.53 second acquisition time. d) Standard error of the mean for the isoprene number density. e) Standard error of the mean for the methane number density.*

Figures 3c-3e compare real-time apodized DCS to traditional DCS for varying apodization window widths and thus resolutions. Figure 3c shows the spectral SNR (see appendix for SNR calculation) as a function of spectral resolution $f_{res}$. For real-time apodized measurements, a linear increase in SNR is observed, which is the same dependence as for varying the fundamental repetition rate $f_r$ of the DCS system [19]. This represents a significant improvement over the $\propto \sqrt{f_{res}/f_r}$ SNR scaling achieved by smoothing of a traditional DCS spectrum through postprocessing apodization.

The standard error on the extracted gas number densities as a function of resolution is shown in Figs. 3d-3e for a 1.53 second acquisition time. For isoprene, a large compound molecule with broad spectral features (see Fig. 2a), it improves $\propto \sqrt{f_r/f_{res}}$ for the real-time apodized data due to the increasing number of measurements within the same time, while it remains flat for traditional DCS. Note that as expected, post-processing apodization, a form of smoothing, improves the spectral SNR (Fig. 3c) but not the number density standard error. For methane, the narrow feature linewidth complicates this simple trend. The standard error first decreases with coarser resolution up to the ~1 GHz width of the methane features, and then plateaus. We attribute this to increased crosstalk between the apodized methane model and the cubic baseline.

These data demonstrate the advantages of a flexible DCS system with real-time apodization. A coarser resolution can be chosen to increase update rate and hence SNR within a given measurement time. On the other hand, one can run the same DCS at fine resolution to measure narrow absorbers with high fidelity, *e.g.*, for decluttering of multiple species or for lineshape characterization.

## 4. Real-time apodization for hyperspectral imaging

Hyperspectral imaging, where one seeks information about the spatial distribution of spectral content in a scene or a sample, is applied to chemical and remote earth sensing, in quality control and medical science [22–25]. Dual-comb spectroscopy, with its broad spectral coverage $\Delta \nu$ and direct mapping of the optical signal into the RF domain is promising for hyperspectral imaging of gases for example. One limitation is that cameras in the near and mid-infrared, where most gases have strong spectral features, have slow frame rates $R$. Matching the RF bandwidth of the optical broad-band DCS signal to this low framerate is challenging, especially for sub-GHz repetition rate combs. For traditional DCS, the maximum hyperspectral movie update rate is given by $R_{hyp} = \Delta f_r \leq f_r R / (2 \Delta \nu)$. For a $\Delta \nu$ =20 THz wide optical signal, a $R$ =1 kHz camera framerate and a $f_r$ =160 MHz DCS system, the maximum hyperspectral frame rate is 0.002 Hz or one image every 500 seconds. Researchers have circumvented this issue by use of high repetition-rate EO-combs and demonstrated hyperspectral update rates of 0.5 Hz to 10 Hz at sub-THz spectral bandwidth [14–18].

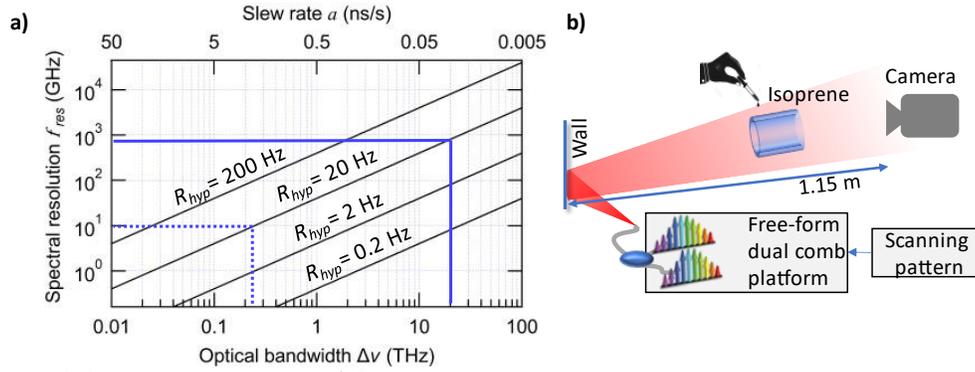

*Figure 4: Hyperspectral imaging. a) Relationship between spectral resolution and maximum optical bandwidth coverage for real-time apodization for a 1 kHz camera frame rate. The diagonal lines correspond to the labeled hyperspectral movie frame rates. In our example, highlighted by the blue lines, a 20 Hz hyperspectral movie update rate is achieved with 20 THz spectral coverage and 822 GHz spectral resolution. Another example, highlighted by blue dashed lines, might be appropriate for narrower absorbers with a spectral resolution of 10 GHz and an optical bandwidth of 250 GHz. b) Hyperspectral imaging setup. The combs are combined in fiber and send into free space from a fiber connector tip. Their diverging light (~3 mW) diffusely back illuminates an open cell, which is filled with liquid isoprene that then evaporates. An InGaAs camera captures the scene.*

Here, we take advantage of the versatility of real-time apodization to trade spectral resolution, update rate and optical bandwidth. The triangular apodized $\tau_{RPD}$ scanning pattern is defined by two parameters, the measurement time per scan, $T_{apod}$, and the slew rate $a$. For a desired update rate $R_{hyp}$ and optical bandwidth $\Delta \nu$, one then sets $a = R / (2\Delta \nu)$ and $T_{apod} = 1 / R_{hyp}$, resulting in an optical resolution of $f_{res} = 1 / (a T_{apod})$. This interdependence is shown in Fig. 4a). Again, this is equivalent to using a higher-repetition rate dual-comb spectrometer with $f_r = f_{res}$. As shown in Fig. 4a, real-time apodization can support both broad spectral coverage and fast hyperspectral update rates independent of frequency comb repetition rate, other than $f_{res} \geq f_r$.

As a demonstration, we select the hyperspectral imaging conditions indicated by the blue solid lines in Figure 4, namely a hyperspectral movie update rate $R_{hyp}$ = 20 Hz and a spectral coverage of $\Delta \nu = 20.6$ THz, resulting in a spectral resolution of $f_{res} = 822$ GHz, suitable for a larger molecular compound. The spectral coverage exceeds previous hyperspectral dual-comb demonstrations by orders of magnitude [14–18].

Figure 5a shows individual frames of a movie acquired over a few seconds that records liquid isoprene drops falling into a glass cell and then evaporating (see Fig. 4b for measurement setup). The raw camera output generates 12-bit monochrome intensity images of the cell illuminated by the dual-comb light. The images are then grouped in sequences of 50, corresponding to a single apodized scan. Each sequence is then Fourier transformed for each pixel yielding a hyperspectral movie at a $R_{hyp} = 20$ Hz update rate. Figure 5b shows raw spectra for a single pixel at different times, where the developing isoprene signature is clearly visible around 184 THz. The spectra of each pixel are first normalized by an empty-cell background spectrum and a third-order polynomial baseline. They are then fitted to an isoprene model [26], yielding isoprene gas time evolution as shown for a single pixel in Figure 5c. After 10 seconds, the isoprene concentration is 28 mol/m$^3$ the equivalent of 50% isoprene at atmospheric conditions in this 8 cm long cell. Figure 5d shows example isoprene concentration images at different times.

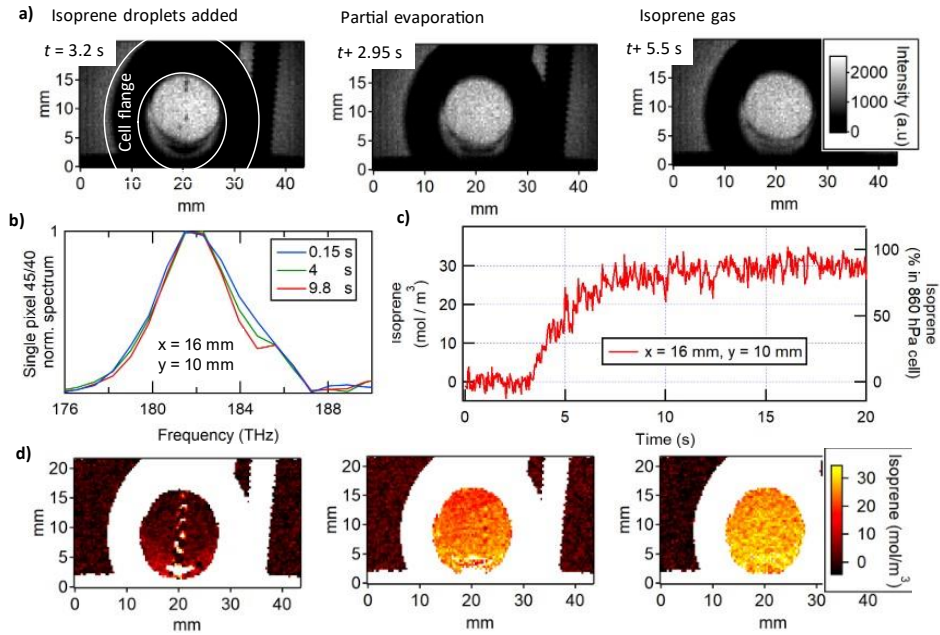

*Figure 5: Hyperspectral imaging with a 128x64 pixel InGaAs NIR camera of isoprene evaporating in a 15 mm wide and 80 mm long cell. a) Raw images from the 1 kHz frame rate movie. Isoprene droplets are added 3.2 s into the measurement and then evaporate over the next few seconds. In the intensity images, the isoprene gas is not visible. b) Example DCS spectra of a single pixel inside the cell at the selected 822 GHz spectral resolution. The time-varying isoprene signature is clearly visible around 184 THz. c) Time evolution of the extracted isoprene gas concentration for the same single pixel over the entire movie duration. d) Example Isoprene concentration images at times indicated in a). The acquisition rate is $R_{hyp} = 20$ Hz. Note that the isoprene droplets are not stationary at this frame rate and cause distortion in the measured spectra of these pixels, resulting in nonsensical extracted isoprene concentrations around the falling droplets at 3.2 seconds. See suppl. material for the full isoprene concentration movie.*

## 5. Conclusions

Real-time apodization expands the measurement capabilities of a dual-comb platform for spectroscopic sensing. It can diversify the applications of a single DCS platform to optimally detect atmospheric trace gases, large molecules, liquids and solids as for example in photo-acoustic measurements [27,28] or muscle tissue signature where broad spectral coverage and fast update rate is needed while coarse spectral resolution is adequate [29]. We show that a free-form dual-comb spectrometer using programmable combs can break free from the traditional preset and fixed repetition rates of the combs used in traditional DCS. We demonstrate and analyze spectral resolution tuning from the high intrinsic 160 MHz, set by the comb repetition rate, to 822 GHz with dead-time free real-time apodization. The spectral SNR increases linearly with the resolution, alike using a higher repetition rate comb source, but this system retains the advantage of a sub-GHz comb source's high-power pulses for spectral broadening and tuning. The programmable real-time apodization enables direct comb hyperspectral imaging at over 10 THz optical bandwidth, orders of magnitude broader than previous demonstrations using high repetition rate frequency combs.

Acknowledgements: We thank T. Bui and B. Washburn for technical comments and acknowledge funding from Rehabilitation Institute of Chicago (Shirley Ryan AbilityLab) subaward No 80335 UCB.Y1 and NIST.

## 7. Appendix

*Phase lock adjustment in free-from dual-comb spectroscopy*

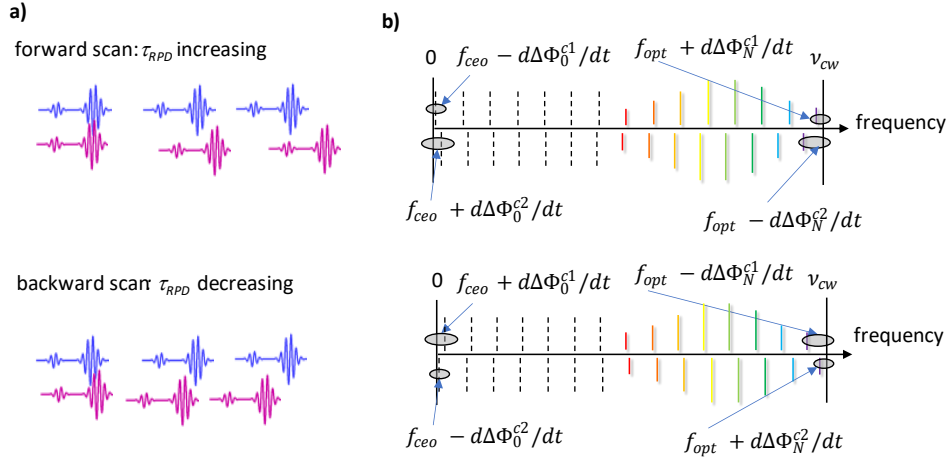

*Figure A1: Schematic overview of time a) and frequency domain b) frequency-domain cartoon for real-time apodized scanning. To initiate scanning of the dual-comb relative pulse delay pulses, the combs are first set to the same repetition rate, followed by phase offsets, with the appropriate signs to the phase locks.*

Figure A1 schematically depicts how the triangular forward and backward scanning over the optical interferogram is implemented. Upper row is a cartoon of how, first, the relative pulse offset is increasing the pink comb pulses walk over the blue comb pulses, and after changing the pulse periods the relative pulse offset is decreasing and the blue pulses walk over the pink pulses (lower row).

The traditional dual-comb sampling, is based on preset difference $\Delta f_r = f_{r1} - f_{r2}$ with acquisition time $1/\Delta f_r$ and pre-set spectral resolution $(f_{r1} + f_{r2})/2 \approx f_r$, which means that the interferogram is always sampled from $-1/(2f_r)$ to $+1/(2f_r)$ with fixed increments in relative pulse offsets of $\Delta f_r / f_r^2$. In order to break these preset relationships, we adapt the phase locking of the combs [12,13], see Fig. A1 b. While phase coherence can be established via different means, here, we lock two comb tooth frequencies, $f_{ceo}$ and $v_{opt}$. The carrier-envelope-offset frequency $f_{ceo} = r_0 f_r$ is stabilized via an $f$-to-$2f$ setup. $N$ is the tooth number nearest to a cw reference laser at $v_{cw}$ = 191.56 THz (6390 cm$^{-1}$), and $v_{opt} = v_{cw} - r_N f_r$ ($r_0$ and $r_N$, $r \in \mathbb{Q}$, $-½ < r < ½$). This approach gives fine control of $f_r = (v_{opt} - f_{ceo})/N = v_{cw}/(N + r_0 + r_N)$ due to the large spacing in frequency between $f_r \sim$ 160 MHz and $v_{opt} \sim$ 191.56 THz and $N\sim1.2\ 10^6$.

After setting both combs to the same repetition rate and establishing full dual-comb pulse overlap, the relative dual-comb pulse delay $\tau_{RPD}$ is changed by programing the control phase $\Delta\theta_0^{c1}, \Delta\theta_0^{c2}, \Delta\theta_N^{c1}, \Delta\theta_N^{c2}$ of the digital $f_0$ and $f_{opt}$ locks acting on both combs (superscript 1 and 2), resulting in a time shift

$$\Delta\tau(t) = \frac{\Delta\theta_0^{c1}(t) + \Delta\theta_0^{c2}(t) - \Delta\theta_N^{c1}(t) - \Delta\theta_N^{c2}(t)}{2\pi N f_r} \qquad (1)$$

We use $\Delta\theta_0^{c1}(t) = -\Delta\theta_N^{c1}(t) = \Delta\theta(t)$, resulting in

$$\Delta\tau(t) = \frac{\Delta\theta(t)}{\pi N f_{rep}} \quad (2)$$

The slew rate at which the relative pulse-offset is changing is $a = \Delta f_r(t)/f_r$ and $\tau_{RPD}$ is swept between $-1/(2f_{res})$ and $+1/(2f_{res})$ where $f_{res} = 1/\tau_{apod} = 1/(aT_{apod})$ is the desired spectral resolution and $T_{apod}$ is the acquisition time for one sweep.

$\Delta f_r(t) = k \cdot d(\Delta\theta(t))/dt$ where $k=2$ indicates that here two actuators (one fast piezo-electric transducer and one current modulation of one comb) are steered, giving a slew rate of ~ 33 ns/s. This slew rate is not a fundamental limit and could be doubled by steering all four locks synchronously [12]. Other limiting factors specific to our setup are actuator throw and the fixed digital bandpass filter center frequency of the digital locks.

When scaling the measured spectrum back to the optical domain, the change to the CEO lock frequency has to be considered and comb teeth overlap is shifted from 0 to $\delta f_{ceo} f_r / \Delta f_r$ where $\delta f_{ceo}$ is the total amount of one or both shifts to the CEO lock frequencies $f_{ceo}$.

For apodization the optical resolution is then $f_{res} = f_r/(T_{apod}\Delta f_r(t)) = 1/\tau_{apod}$.

There is a small deadtime in real-time apodization caused by the finite bandwidth of the frequency comb actuators. Here, the $f_{CEO}$ lock frequency is modulated via the pump diode current at an amplitude of +/- 3.125 MHz, and it takes about 50 µs for the $f_{CEO}$ to settle after reversal of comb pulse period (see Fig. A2). For the apodized data presented in this manuscript, the first 50 microsecond of each sweep are not used and present effectively dead time. This could be improved by either using faster actuators or digitally correcting for the frequency transient immediately following jumps by calculating the effective pulse offset time of each interferogram point and resampling on an equidistant grid [8].

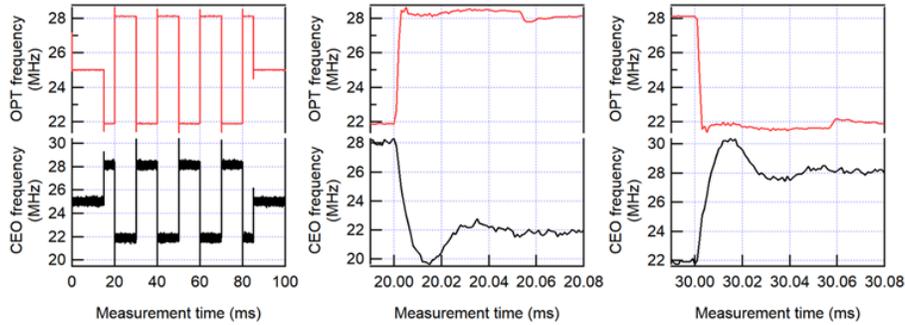

Figure A2: Actuator response for CEO and optical (OPT) instantaneous lock frequency beat for a +/- 3.125 MHz jump around the 25 MHz set value. The CEO lock, modulated via the oscillator pump diode current, has a slower response and takes about 50 microseconds to fully settle. The optical lock, modulated via a fast intracavity PZT, is faster and is mostly settled after 5 microseconds. The response is antisymmetric in both directions.

*Spectral SNR calculation*

Figure A3 describes how the spectral SNR shown in the main body Fig. 3d is calculated.

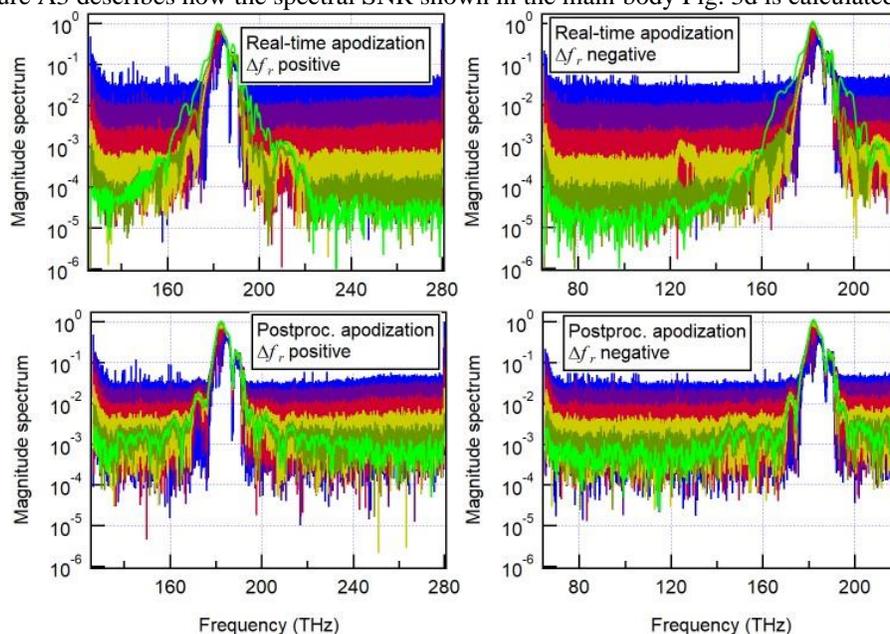

*Figure A3: Acquired spectra averaged over a measurement time of 1.53 second for real-time apodization (top panels) and postprocessing apodization of fully resolved spectra (bottom panels). Each trace corresponds to a specific resolution, ranging from 160 MHz (blue, highest noise level) to 241 GHz (bright green, lowest noise level). To obtain the averaged magnitude spectrum for a given resolution, the complex spectra are phase corrected, coadded and transformed to magnitude. Because the pulse period of the combs changes between consecutive acquisitions, the optical span covered is different for 'positive' and 'negative' scan direction, as shown in left versus right panels. The signal is defined as peak magnitude, the noise as average magnitude between 250 THz and 260 THz for positive scan direction respectively between 100 THz and 110 THz for negative.*

*Spectral fitting*

This section describes in detail how isoprene and methane number densities are extracted from the raw measured spectra through an isoprene and methane cell. The experimental setup is shown in Figure 2 a of the main manuscript.

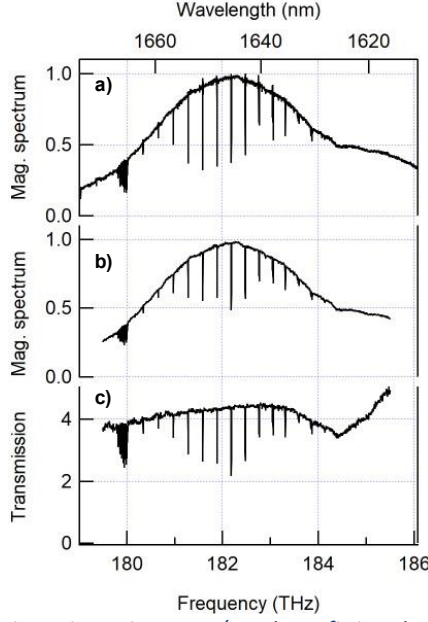

*Figure A4: Spectrum preparation prior to isoprene / methane fitting shown for 2.57 GHz resolution. a) Raw spectrum apodized to 2.57 GHz showing spectral signature of the DCS (overall baseline), methane (narrow lines) and isoprene (broad feature, most prominent around 184.5 THz). b) Same spectrum spanning 179.5 THz to 185.5 THz after applying a Hann window. c) Spectrum from b) normalized with a reference spectrum acquired simultaneously on a separate acquisition channel (see Figure 2a).*

Only spectral data between 179.5 THz and 185.5 THz is used for fitting. After cropping the spectrum, the raw signal spectrum shown in Fig.A4 a is then inverse Fourier transformed, multiplied with a Hann window, and Fourier transformed to obtain the windowed spectrum shown in Fig.A4 b. This is done to reduce the ringing surrounding the narrow methane feature which increases fit robustness. The windowed signal spectrum is then divided by a reference spectrum which was previously smoothed to a 20 GHz resolution to avoid adding excess noise, Fig.A4c.

This normalized transmission spectrum is then fit with an iterative nonlinear least-squares Levenberg-Marquardt algorithm to the model

$$(c_0 + c_1 p + c_2 p^2 + c_3 p^3) * \exp(-c_4 * iso_{PNNL} - c_5 * CH_{4_{Hitran}})$$

where $c_x$ are the floating fit parameters: $c_0, c_1, c_2, c_3$ are the cubic polynomial baseline parameters, $c_4$ is the path integrated isoprene concentration at 1013.25 hPa and 296 K and $c_5$ is the path integrated methane concentration at 60 hPa and 296 K. Finally, using the known gas cell lengths $l_{isocell} = 0.06$ m and $l_{ch4cell} = 0.75$ m, $c_4$ and $c_5$ are converted to the number densities $ND$ reported in the main body via the ideal gas law:

$$ND_{iso} = c_4/l_{isocell} * \frac{101325}{296 * R}, \quad ND_{CH4} = c_5/l_{ch4cell} * \frac{6000}{296 * R}$$

where $R = 8.31446261815324$ is the molar gas constant.

The isoprene spectrum $iso_{PNNL}$ is taken from the PNNL data base [21], which provides a measured reference spectrum at 296 K and 101325 Pa. The methane spectrum $CH_{4_{Hitran}}$ is calculated using Hitran 2008 with a Voigt lineshape. First, both the methane concentration and the methane cell ambient pressure were fit with the 160 MHz resolution spectra, resulting in a fitted ambient methane gas cell pressure of 6000 Pa and a methane concentration of 40%. Then, the methane model $CH_{4Hitran}$ was calculated for these conditions and scaled with a fit concentration coefficient $c_5$ for all fits. Finally, both $iso_{PNNL}$ and $CH_{4Hitran}$ are apodized to the resolution of the measurement and a Hann window is applied to the model.

Figure A5 gives an overview of the good fit qualities at all measured spectral resolution applying the adequate apodization window to the model. At $f_{res}$ = 241 GHz, $\tau_{RPD}$ is scanned symmetrically from -2.07 ps to +2.07 ps. The first methane recurrence is not covered anymore, the spectral resolution is close to the periodic line spacing of the lines, and the fit result for methane is strongly influenced by baseline crosstalk. Isoprene on the other side is still captured reliably.

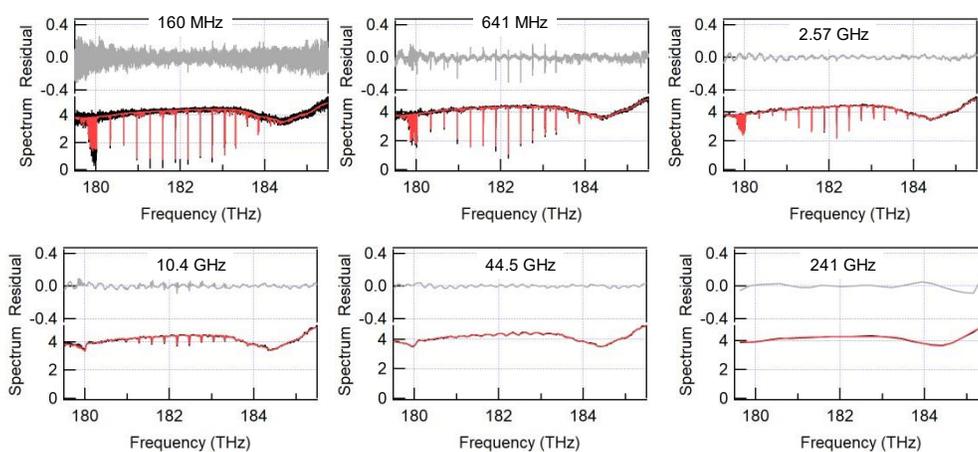

*Figure A5: Normalized transmission spectra of methane and isoprene in black for the spectral resolution as labeled in the panels. Red: transmission model based on Hitran 2008 (CH₄) [30] and PNNL (isoprene) [21] databases, using a Hann apodization window, for a single scan. Top (grey), residuals to the fit, averaged over the entire measurement time of 1.53 second to highlight structure not captured by the fit model.*